\documentclass{article}
\usepackage{spconf,amsmath,graphicx}
\usepackage{amsfonts}
\usepackage{amsmath}
\usepackage{subfigure}
\usepackage{mathrsfs}
\usepackage{multirow}
\usepackage{bm}
\usepackage{verbatim}
\usepackage[normalem]{ulem}
\usepackage{float}

\usepackage{mathtools,lipsum,cuted}
\usepackage{tikz,pgfplots,pgfplotstable}
\usepgfplotslibrary{groupplots,fillbetween,colorbrewer,statistics}
\usetikzlibrary{intersections,calc,arrows,matrix,spy,pgfplots.statistics, pgfplots.colorbrewer}
\usepackage{color}
\definecolor{darkred}{rgb}{0.7,0,0}
\definecolor{darkgreen}{rgb}{0,0.5,0}
\definecolor{darkblue}{rgb}{0,0,0.7}
\definecolor{SkyBlue}{rgb}{0.53, 0.81, 0.92}
\pgfplotsset{compat=1.5.1, cycle list/Set1-3}
\usepackage{xfrac}
\usepackage[colorlinks]{hyperref}
\usepackage[short]{optidef}
\usepackage{bbm}

\usepackage{tikz}
\usepackage{float}

\newsavebox{\tempbox}

\allowdisplaybreaks
\usepackage[]{caption} 
\usepackage{graphicx}
\usepackage[export]{adjustbox}
\usepackage{caption}

\usepackage{enumitem}

\usepackage[ruled]{algorithm}
\usepackage{algpseudocode,algorithmicx}

\title{MSR-GAN: Multi-Segment Reconstruction via Adversarial Learning}
\name{Mona Zehni, Zhizhen Zhao}
\address{Department of ECE and CSL, University of Illinois at Urbana-Champaign}

\begin{document}
\savebox{\tempbox}{\begin{tabular}{@{}r@{}l@{\space}}
&\scriptsize{SNR}\\ \scriptsize{$M$}
\end{tabular}}

\newcommand{\remove}[1]{}

\ninept
\maketitle

\begin{abstract}
Multi-segment reconstruction (MSR) is the problem of estimating a signal given noisy partial observations. Here each observation corresponds to a randomly located segment of the signal. While previous works address this problem using template or moment-matching, in this paper we address MSR from an unsupervised adversarial learning standpoint, named MSR-GAN. We formulate MSR as a \textit{distribution} matching problem where the goal is to recover the signal and the probability distribution of the segments such that the distribution of the generated measurements following a known forward model is close to the real observations. This is achieved once a min-max optimization involving a generator-discriminator pair is solved. MSR-GAN is mainly inspired by CryoGAN~\cite{cryogan}. However, in MSR-GAN we no longer assume the probability distribution of the latent variables, i.e. segment locations, is given and seek to recover it alongside the unknown signal. For this purpose, we show that the loss at the generator side originally is non-differentiable with respect to the segment distribution. Thus, we propose to approximate it using Gumbel-Softmax reparametrization trick. Our proposed solution is generalizable to a wide range of inverse problems. Our simulation results and comparison with various baselines verify the potential of our approach in different settings.
\end{abstract}
\keywords{Multi-segment reconstruction, adversarial learning, unsupervised learning, Gumbel-Softmax approximation, categorical distribution.}
\section{Introduction}
\label{sec:intro}
\begin{figure}
    \centering
    \includegraphics[width=1 \linewidth]{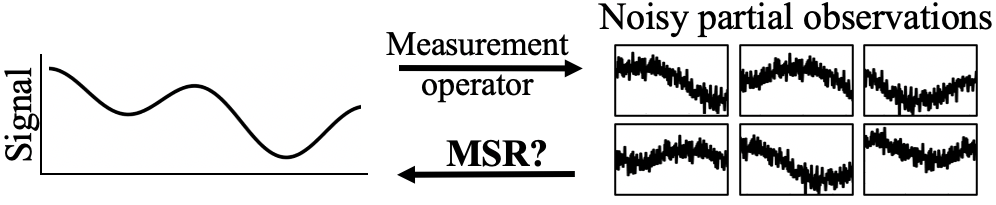}
    \caption{Multi-segment reconstruction (MSR) problem.}
    \label{fig:msr}
    \vspace{-20pt}
\end{figure}

\vspace{-6pt}
The problem of recovering a signal from a set of noisy partial observations appear in a wide range of applications including genomic sequence assembly~\cite{Motahari2013}, puzzle solving\cite{Paikin2015}, tomographic reconstruction~\cite{Willemink2018} and cryo-electron microscopy (Cryo-EM)~\cite{Barnett2016, Punjani2017}, to name a few. In this paper, we focus on multi-segment reconstruction (MSR)~\cite{zehni_msr}, where the unknown is a 1D sequence and the measurements are noisy randomly located partial observations (segments) of this sequence. A schematic illustration of MSR is provided in Fig.~\ref{fig:msr}. MSR is a general form of multi-reference alignment (MRA)~\cite{Bendory2017} problem in which the measurements are noisy randomly shifted versions of the signal. While in MRA the length of each measurement is the same as the signal, in MSR the measurements can be shorter.

Current efforts devoted to MSR is studied in two broad categories, 1) alignment-based, 2) alignment-free. In one form of alignment-based methods, the segment location corresponding to each observation is estimated. Then, the observations are aligned accordingly and averaged. While these methods have low computational and sample complexity, low signal-to-noise ratio (SNR) of the observations adversely affect their performance. Examples of alignment-based methods applied to MRA and tomographic reconstruction are found in~\cite{chen2016, basu2000_noise}. In other forms of alignment based methods, the segment locations and the 1D sequence are jointly updated using alternating steps. An example would be the maximum likelihood formulation of MSR, solved using expectation-maximization (EM). Despite the robustness of EM to different noise regimes, it suffers from high computational complexity. This is due to the complexity of the E-step, requiring a whole pass through the measurements at every iteration. This is significantly time-consuming, especially in the presence of large number of observations.

Alignment-free solutions specifically designed for MRA side-step the estimation of the random shifts by introducing a set of invariant features. These features constitute the moments of the signal and are estimated from the measurements. The signal is then estimated from the features via an optimization-based framework~\cite{Bendory2017, Boumal2017}, tensor decomposition~\cite{bandeira2020optimal,Kolda2009} using Jennrich's algorithm~\cite{harshman70} or spectral decomposition~\cite{chen2018,abbe2017sample}. As these works are specialized for MRA, they do not address the challenges associated with MSR, such as observing only shorter segments of the signal. In~\cite{zehni_msr}, we showed how for MSR, we can estimate the invariant features from the measurements and how the recovery of the signal is tied to the segment length. Compared to alignment-based solutions, in alignment-free methods, we only have one pass through the measurements to estimate the features, thus computationally more efficient. The estimated features then serve as a compact representation of the measurements which are functions of the unknown signal and segment location distribution. 

In this paper, we propose an alignment-free adversarial learning based method for MSR. Our goal is to find the unknown 1D signal and the distribution of the segment locations such that the measurements generated from the estimated signal match the real measurements in a distribution sense. Therefore, we train a generator discriminator pair, where the discriminator tries to distinguish between the measurements output by the generator and the real ones. Our approach is inspired by CryoGAN~\cite{cryogan} in which the goal is to reconstruct a 3D structure given 2D noisy projection images from unknown projection views. Unlike CryoGAN, we assume the distribution of the latent variables, i.e. the segment locations in MSR, is unknown and we seek to recover it alongside the signal. For this purpose, we modify the loss at the generator side using Gumbel-Softmax approximation of categorical distribution, to accommodate gradient-based updates of the segment location distribution. Our simulation results and comparison with several baselines confirm the feasibility of our approach in various segment length and noise regimes. Our code is available at \url{https://github.com/MonaZI/MSR-GAN}.
\vspace{-10pt}

\section{System Model}
\label{sec:model}
\vspace{-6pt}
\begin{figure}
    \centering
    \includegraphics[width=1 \linewidth]{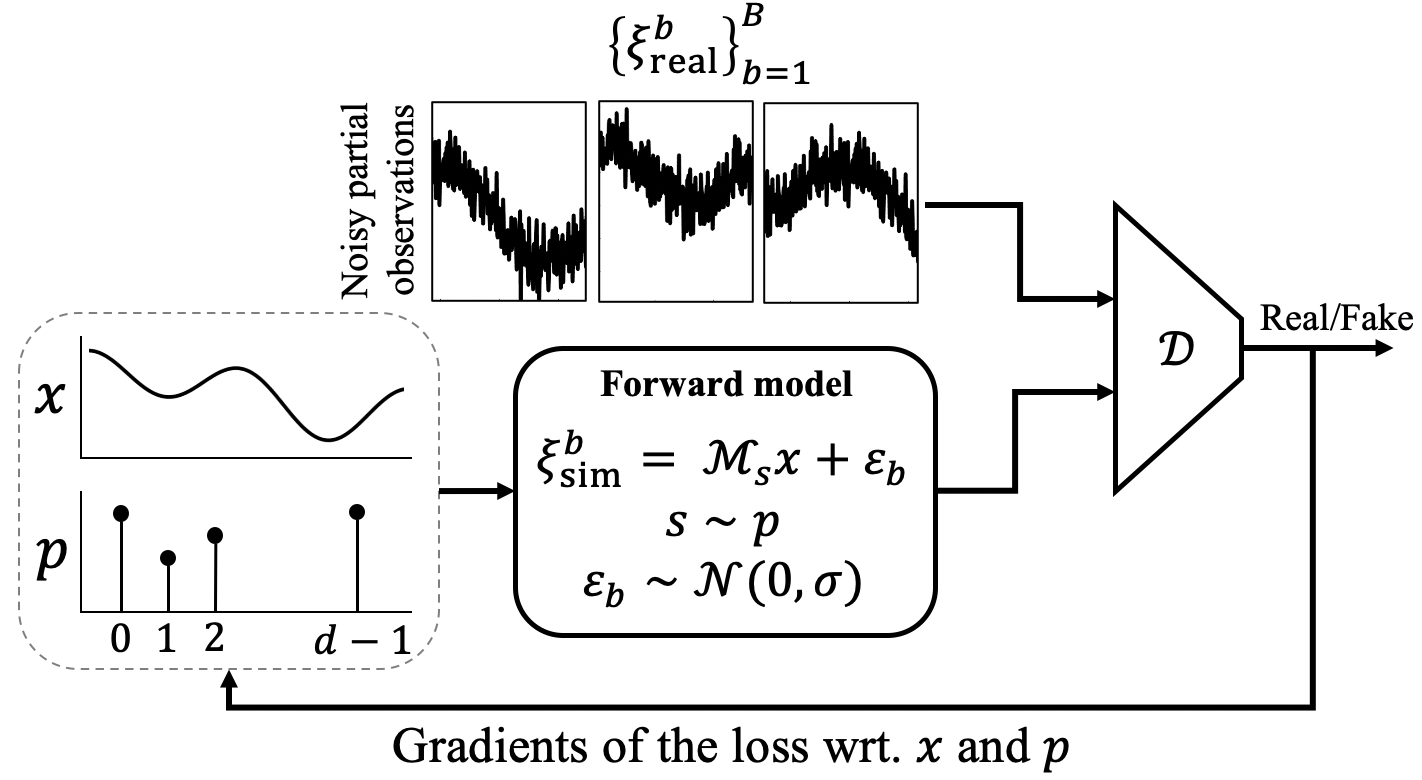}
    \caption{An illustration of MSR-GAN pipeline.} 
    \label{fig:msrgan_pipeline}
\end{figure}

We consider the following observation model,
\begin{equation}
\label{eq:obs_model}
\xi_j = \mathcal{M}_{s_j} x + \varepsilon_j, \quad j \in \{1,2,...,N\}
\end{equation}
where $x \in \mathbb{R}^d$ is the underlying signal and $\xi_j \in \mathbb{R}^m$, $m \leq d$ is the $j$-th observation. We often refer to $m$ as the segment length. The cyclic masking operator $\mathcal{M}_{s}$ captures $m$ consecutive entries of $x$ starting from index $s$. In other words, $\mathcal{M}_s: \mathbb{R}^d \rightarrow \mathbb{R}^m$ and $\left(\mathcal{M}_{s}x\right)[n] = x[n+s \, \textrm{mod} \, d]$. We also assume the segment location $s \in \{0,1,...,d-1\}$ to be unknown and randomly drawn from a categorical distribution with $p$ as its probability mass function (PMF) where ${P}\{s=s_j\} = p[s_j]$. 
In addition, the randomly located segment of the signal is contaminated by additive white Gaussian noise $\varepsilon_j$ with zero mean and covariance $\sigma^2 I_m$ ($I_m$ denoting the identity matrix with size $m \times m$). Our goal here is to recover $x$ and $p$ given the noisy partial observations $\{\xi_j\}_{j=1}^N$. 

Note that the distribution of the observations depends on both the signal $x$ and the distribution of the segment locations $p$. Thus, it is possible to estimate $x$ and $p$ by matching the distribution of the observations generated by $x$ and $p$ following~\eqref{eq:obs_model} to the real measurements.
\vspace{-10pt}

\begin{figure*}[h]
\centering
	\begin{tikzpicture}
		\begin{groupplot}[group style={group size= 4 by 3,                      
    				horizontal sep=0.6cm, vertical sep=1.cm},     
	            	 legend pos= south west,        
					 legend style={at={(0.45,0.5)}, legend cell align=left, draw=none, fill=none},
					 grid=both,                         
    				 height=3.5cm,width=5.4cm,
    				 xmin=1,xmax=64,
					 ymin=-0.1,ymax=1, 
					 ylabel near ticks, xlabel near ticks, xlabel style={align=center,text width=3cm},
					 yticklabel style = {font=\tiny,yshift=0.5ex},
					 xticklabel style = {font=\tiny,xshift=0.5ex},] 
		\nextgroupplot[title={Known PMF},xticklabels={,,},xlabel={\textcolor{darkblue}{$\textrm{Rel-Error}=0.0008$} \textcolor{darkred}{$\textrm{Rel-Error}=0.0130$}},ymin=-1.2,ymax=1.2, x label style={at={(axis description cs:0.5,0)},anchor=north}] 	
			 \addplot[darkblue,very thick] table[x=x, y=y, col sep=comma]{figs/viz_signals_dat_new/0_est_known_sin_new.dat};
			 \addplot[darkred,very thick] table[x=x, y=y, col sep=comma]{figs/viz_signals_dat_new/1_est_known_sin_new.dat};
			 \addplot[darkgreen,very thick,dashdotted] table[x=x, y=y, col sep=comma]{figs/viz_signals_dat/0_gt_known_sin.dat};
		\legend{{\tiny $\textrm{SNR}=\infty$ },{\tiny $\textrm{SNR}=1$ } ,{\tiny GT}};
		\nextgroupplot[title={Fixed PMF with Unif.}, xticklabels={,,},xlabel={\textcolor{darkblue}{$\textrm{Rel-Error}=0.013$} \textcolor{darkred}{$\textrm{Rel-Error}=0.1251$}},ymin=-1.2,ymax=1.2, x label style={at={(axis description cs:0.5,0)},anchor=north}] 	
			\addplot[darkblue,very thick] table[x=x, y=y, col sep=comma]{figs/viz_signals_dat_new/0_est_fixed_sin_new.dat};
			\addplot[darkred,very thick] table[x=x, y=y, col sep=comma]{figs/viz_signals_dat_new/1_est_fixed_sin_new.dat};
			\addplot[darkgreen,very thick,dashdotted] table[x=x, y=y, col sep=comma]{figs/viz_signals_dat/0_gt_fixed_sin.dat};
		\nextgroupplot[title={Unknown PMF}, xticklabels={,,},xlabel={\textcolor{darkblue}{$\textrm{Rel-Error}=0.0022$} \textcolor{darkred}{$\textrm{Rel-Error}=0.0085$}},ymin=-1.2,ymax=1.2, x label style={at={(axis description cs:0.5,0)},anchor=north}] 	
			 \addplot[darkblue,very thick] table[x=x, y=y, col sep=comma]{figs/viz_signals_dat_new/0_est_unknown_sin_new.dat};	
			 \addplot[darkred,very thick] table[x=x, y=y, col sep=comma]{figs/viz_signals_dat_new/1_est_unknown_sin_new.dat};	
			 \addplot[darkgreen,very thick,dashdotted] table[x=x, y=y, col sep=comma]{figs/viz_signals_dat/0_gt_unknown_sin.dat};
	    \nextgroupplot[title={MSR-GAN est. PMF vs GT},xticklabels={,,},xlabel={\textcolor{darkblue}{$\textrm{TV}=0.023$} \textcolor{darkred}{$\textrm{TV}=0.059$}}, ymin=0.001,ymax=0.035, x label style={at={(axis description cs:0.5,0)},anchor=north}] 	
			 \addplot[darkblue,very thick] table[x=x, y=y, col sep=comma]{figs/viz_signals_dat_new/0_est_pdf_sin_new.dat};	
			 \addplot[darkred,very thick] table[x=x, y=y, col sep=comma]{figs/viz_signals_dat_new/1_est_pdf_sin_new.dat};	
			 \addplot[darkgreen,very thick,dashdotted] table[x=x, y=y, col sep=comma]{figs/viz_signals_dat/0_gt_pdf_sin.dat};
		\nextgroupplot[xticklabels={,,},xticklabels={,,},xlabel={\textcolor{darkblue}{$\textrm{Rel-Error}=0.0003$} \textcolor{darkred}{$\textrm{Rel-Error}=0.0320$}},ymin=-0.2,ymax=1.2, x label style={at={(axis description cs:0.5,0)},anchor=north}] 	
			 \addplot[darkblue,very thick] table[x=x, y=y, col sep=comma]{figs/viz_signals_dat_new/0_est_known_tri_new.dat};
			 \addplot[darkred,very thick] table[x=x, y=y, col sep=comma]{figs/viz_signals_dat_new/1_est_known_tri_new.dat};
			 \addplot[darkgreen,very thick,dashdotted] table[x=x, y=y, col sep=comma]{figs/viz_signals_dat/0_gt_known_tri.dat};
		\nextgroupplot[xticklabels={,,},xticklabels={,,},xlabel={\textcolor{darkblue}{$\textrm{Rel-Error}=0.04$} \textcolor{darkred}{$\textrm{Rel-Error}=0.0858$}},ymin=-0.2,ymax=1.2, x label style={at={(axis description cs:0.5,0)},anchor=north}] 	
			\addplot[darkblue,very thick] table[x=x, y=y, col sep=comma]{figs/viz_signals_dat_new/0_est_fixed_tri_new.dat};
			\addplot[darkred,very thick] table[x=x, y=y, col sep=comma]{figs/viz_signals_dat_new/1_est_fixed_tri_new.dat};
			\addplot[darkgreen,very thick,dashdotted] table[x=x, y=y, col sep=comma]{figs/viz_signals_dat/0_gt_fixed_tri.dat};
		\nextgroupplot[xticklabels={,,},xlabel={\textcolor{darkblue}{$\textrm{Rel-Error}=0.0006$} \textcolor{darkred}{$\textrm{Rel-Error}=0.0230$}}, x label style={at={(axis description cs:0.5,0)},anchor=north}, ymin=-0.2,ymax=1.2] 	
			 \addplot[darkblue,very thick] table[x=x, y=y, col sep=comma]{figs/viz_signals_dat_new/0_est_unknown_tri_new.dat};
			 \addplot[darkred,very thick] table[x=x, y=y, col sep=comma]{figs/viz_signals_dat_new/1_est_unknown_tri_new.dat};
			 \addplot[darkgreen,very thick,dashdotted] table[x=x, y=y, col sep=comma]{figs/viz_signals_dat/0_gt_unknown_tri.dat};
	    \nextgroupplot[xticklabels={,,},xlabel={\textcolor{darkblue}{$\textrm{TV}=0.035$} \textcolor{darkred}{$\textrm{TV}=0.046$}},ymin=0.003,ymax=0.032, x label style={at={(axis description cs:0.5,0)},anchor=north}] 	
			 \addplot[darkblue,very thick] table[x=x, y=y, col sep=comma]{figs/viz_signals_dat_new/0_est_pdf_tri_new.dat};	
			 \addplot[darkred,very thick] table[x=x, y=y, col sep=comma]{figs/viz_signals_dat_new/1_est_pdf_tri_new.dat};	
			 \addplot[darkgreen,very thick,dashdotted] table[x=x, y=y, col sep=comma]{figs/viz_signals_dat_new/0_gt_pdf_tri_new.dat};
		\nextgroupplot[xticklabels={,,},xticklabels={,,},xlabel={\textcolor{darkblue}{$\textrm{Rel-Error}=0.0016$} \textcolor{darkred}{$\textrm{Rel-Error}=0.0201$}},ymin=-0.6,ymax=0.6, x label style={at={(axis description cs:0.5,0)},anchor=north}] 	
			 \addplot[darkblue,very thick] table[x=x, y=y, col sep=comma]{figs/viz_signals_dat_new/0_est_known_random_new.dat};
			 \addplot[darkred,very thick] table[x=x, y=y, col sep=comma]{figs/viz_signals_dat_new/1_est_known_random_new.dat};
			 \addplot[darkgreen,very thick,dashdotted] table[x=x, y=y, col sep=comma]{figs/viz_signals_dat/0_gt_known_rand.dat};
		\nextgroupplot[xticklabels={,,},xticklabels={,,},xlabel={\textcolor{darkblue}{$\textrm{Rel-Error}=0.0073$} \textcolor{darkred}{$\textrm{Rel-Error}=0.0372$}},ymin=-0.6,ymax=0.6, x label style={at={(axis description cs:0.5,0)},anchor=north}] 	
			\addplot[darkblue,very thick] table[x=x, y=y, col sep=comma]{figs/viz_signals_dat_new/0_est_fixed_random_new.dat};
			\addplot[darkred,very thick] table[x=x, y=y, col sep=comma]{figs/viz_signals_dat_new/1_est_fixed_random_new.dat};
			\addplot[darkgreen,very thick,dashdotted] table[x=x, y=y, col sep=comma]{figs/viz_signals_dat/0_gt_fixed_rand.dat};
		\nextgroupplot[xticklabels={,,},xlabel={\textcolor{darkblue}{$\textrm{Rel-Error}=0.0003$} \textcolor{darkred}{$\textrm{Rel-Error}=0.0205$}},ymin=-0.6,ymax=0.6, x label style={at={(axis description cs:0.5,0)},anchor=north}] 	
			 \addplot[darkblue,very thick] table[x=x, y=y, col sep=comma]{figs/viz_signals_dat_new/0_est_unknown_random_new.dat};
			 \addplot[darkred,very thick] table[x=x, y=y, col sep=comma]{figs/viz_signals_dat_new/1_est_unknown_random_new.dat};
			 \addplot[darkgreen,very thick,dashdotted] table[x=x, y=y, col sep=comma]{figs/viz_signals_dat/0_gt_unknown_rand.dat};
	    \nextgroupplot[xticklabels={,,},xlabel={\textcolor{darkblue}{$\textrm{TV}=0.030$} \textcolor{darkred}{$\textrm{TV}=0.076$}}, ymin=0.004,ymax=0.03, x label style={at={(axis description cs:0.5,0)},anchor=north}] 	
			 \addplot[darkblue,very thick] table[x=x, y=y, col sep=comma]{figs/viz_signals_dat_new/0_est_pdf_random_new.dat};
			 \addplot[darkred,very thick] table[x=x, y=y, col sep=comma]{figs/viz_signals_dat_new/1_est_pdf_random_new.dat};
			 \addplot[darkgreen,very thick,dashdotted] table[x=x, y=y, col sep=comma]{figs/viz_signals_dat_new/0_gt_pdf_random_new.dat};
		\end{groupplot}
	\end{tikzpicture}
\caption{Comparison between MSR-GAN in different noise regimes for 1) known PMF (first column), 2) unknown PMF but fixed with uniform distribution during training (second column), 3) unkown PMF and recovered during training (third column). The last column plots the ground truth PMF (green dashed curve) alongside the estimated PMFs from MSR-GAN (the same experiment as the third column) in blue and red. Each row corresponds to different signals and PMFs. The relative error of the reconstruction for $\textrm{SNR}=\infty$ and $\textrm{SNR}=1$ is written in blue ($\textrm{SNR}=\infty$) and red ($\textrm{SNR}=1$) underneath each subplot. For all experiments in this figure we are using the same architecture for the discriminator with $\ell=100$ and the number of measurements is $N=5 \times 10^4$.}
\label{fig:signal_viz_results}
\end{figure*}
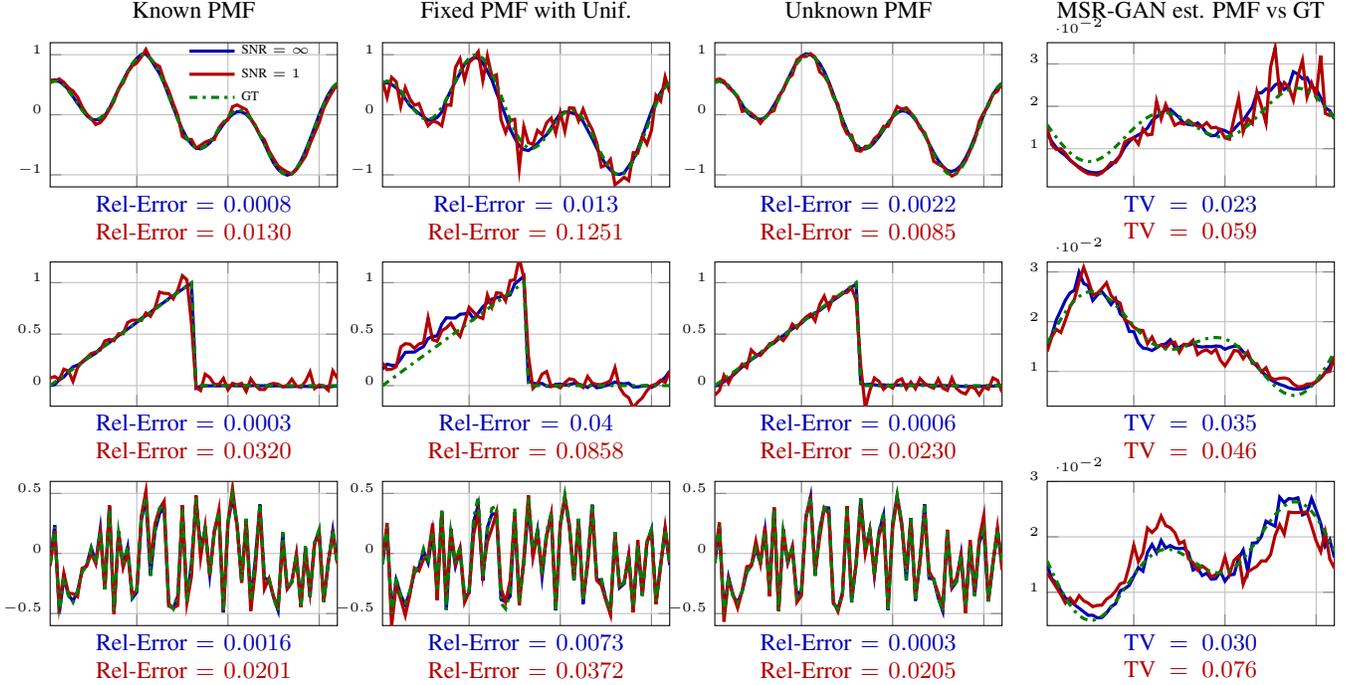

\section{Method}
\label{sec:method}
\vspace{-6pt}
We use an unsupervised adversarial learning approach to solve MSR. Our method is unsupervised as it only relies on the given observations and does not use large paired datasets for training. Similar to~\cite{cryogan}, our method aims to find $x$ and $p$ such that the distribution of the partial noisy measurements generated from~\eqref{eq:obs_model} matches the real measurements $\{\xi^j_{\textrm{real}}\}_{j=1}^N$. To this end, we use a generative adversarial network (GAN)~\cite{NIPS2014_5423}. Unlike common GAN models, we use the known forward model in~\eqref{eq:obs_model} to map the signal and segment distribution to the measurements. Thus, the generator acts upon $x$ and $p$ and simulates noisy measurements $\{\xi^j_{\textrm{sim}}\}_{j=1}^M$. The discriminator's task is then to distinguish between the real and fake measurements from the generator. An illustration of MSR-GAN is provided in Fig.~\ref{fig:msrgan_pipeline}. Here we use Wasserstein GAN~\cite{wgan} with gradient penalty (WGAN-GP)~\cite{wgangp}, to benefit from its favorable convergence behaviour. In WGAN, the output of the discriminator is a score, where the more the input resembles $\xi_{\textrm{real}}$, the higher the score. The min-max formulation of the problem is:
{\small
\begin{align}
    &\widehat{x}, \widehat{p} = \arg \min_{x, p} \max_{\phi} \mathcal{L}(\phi, x, p) \label{eq:minmax}
\end{align}}%
\vspace{-20pt}
\small{\begin{align}
    \mathcal{L}(\phi, x, p) &= \sum\limits_{b=1}^{B} \mathcal{D}_{\phi} (\xi^b_{\textrm{real}}) - \mathcal{D}_{\phi} (\xi^b_{\textrm{sim}})  -
    \lambda \, \textrm{GP}(\xi^b_{\textrm{int}}) \label{eq:loss_function} \\
    \textrm{GP}(\xi^b_{\textrm{int}}) &= \left( \Vert \nabla_{\xi} \mathcal{D}_{\phi}(\xi^b_{\textrm{int}}) \Vert -1 \right)^2 
\end{align}}%
where $\mathcal{L}$ denotes the loss which is a function of the discriminator's parameters $\phi$, the signal and the PMF. Also, $B$ is the batch size, $\mathcal{D}_{\phi}$ denotes the discriminator parameterized by $\phi$ and $\xi_{\textrm{sim}} = \mathcal{M}_s x + \varepsilon$, $s \sim p$ and $\varepsilon \sim \mathcal{N}(0, \sigma^2 I_m)$. The weight of the gradient penalty term (GP) is $\lambda$ and $\xi_{\textrm{int}}$ is a sample generated by linear interpolation between a real and simulated measurement, i.e. $\xi_{\textrm{int}} = \alpha \, \xi_{\textrm{real}} + (1-\alpha) \, \xi_{\textrm{sim}}$, $\alpha \sim \textrm{Unif(0, 1)}$.
To solve the min-max optimization in~\eqref{eq:minmax}, following common practice, we take alternating steps to update the discriminator's parameters $\phi$ and the generator, i.e. $x$ and $p$, using their gradients.

 \begin{algorithm}[t]
   \caption{MSR-GAN}\label{alg:msrgan}
     \textbf{Require:} $\alpha_\phi$, $\alpha_x$, $\alpha_p$: learning rates for the discriminator, the image and projection angle distribution. $\lambda$: gradient penalty weight. $n_{\textrm{disc}}$: the number of iterations of the discriminator (critic) per generator iteration. \\
     \textbf{Require:} Initialize $x$ randomly and $p$ with a uniform distribution, i.e. $p^0[s] = 1/d$.\\
     \textbf{Output:} Estimates $I$ and $p$ given $\{\xi^j_{\textrm{real}}\}_{j=1}^N$. 
   \begin{algorithmic}[1]
     \While{not converged}
     \For{$t=0,...,n_{\textrm{disc}-1}$}
     \State Sample a batch from real data, $\{\xi^b_{\textrm{real}}\}_{b=1}^B$
     \State Sample a batch of simulated measurements using estimated signal and PMF, i.e. $\{\xi^b_{\textrm{sim}}\}_{b=1}^B$ where $\xi^b_{\textrm{sim}} = \mathcal{M}_{s} x + \varepsilon_b$, $\varepsilon_b \sim \mathcal{N}(0, \sigma I_m)$
     \State Generate interpolated samples $\{\xi^b_{\textrm{int}}\}_{b=1}^B$, $\xi^b_{\textrm{int}} = \alpha \, \xi^b_{\textrm{real}} + (1-\alpha) \, \xi^b_{\textrm{sim}}$ with $\alpha \sim \textrm{Unif}(0, 1)$
     \State Update the discriminator using gradient ascent steps with,
     \vspace{-10pt}
     {\small
     \begin{align}
     \nabla_{\phi} \mathcal{L}_{D}(\phi) = \nabla_{\phi} \left( \sum\limits_{b=1}^{B} \mathcal{D}_{\phi} (\xi^b_{\textrm{real}}) - \mathcal{D}_{\phi} (\xi^b_{\textrm{sim}}) +
    \lambda \textrm{GP}(\xi^b_{\textrm{int}}) \right) \nonumber
    \end{align}}%
    \vspace{-15pt}
    \EndFor
     \State Sample a batch of $\{q_{i, b}\}_{b=1}^B$ using~\eqref{eq:gumbel_approx}
     \State Update $x$ and $p$ using gradient descent steps with the following gradients,
     \vspace{-10pt}
     {\small
     \begin{align}
         \nabla_{x, p} \mathcal{L}_G(x, p) &= \nabla_{x, p} \left(- \sum\limits_{b=1}^{B} \sum\limits_{s=0}^{d-1} q_{i, b} \mathcal{D}_{\phi} (\mathcal{M}_s x + \varepsilon_b) \right) \nonumber
     \end{align}}%
     \vspace{-10pt}
     \EndWhile
   \end{algorithmic}
 \end{algorithm}

To update $p$, we need to take gradients of~\eqref{eq:loss_function} with respect to $p$. 
However, this loss function is related to $p$ through a sampling operator which is non-differentiable (we are sampling the segment locations  based on the $p$ distribution). This would be problematic at the generator update steps. Therefore, it is crucial to devise a way to have a meaningful gradient with respect to $p$. First, let us take a closer look at the loss function that is minimized at the generator side:
\vspace{-5pt}
\begin{align}
    \mathcal{L}_G(x, p)\! &=\! -\! \sum\limits_{b=1}^B \mathcal{D}_{\phi}(\mathcal{M}_{s_b} x + \varepsilon_b), \, s\! \sim \! p, \, \varepsilon \! \sim \! \mathcal{N}(0, \sigma^2 I_m) \\
    & = - \sum\limits_{b=1}^B \sum\limits_{s=0}^{d-1} \delta(s - s_b) \mathcal{D}_{\phi}(\mathcal{M}_{s} x + \varepsilon_b)
\end{align}
where $\delta$ is the Kronecker delta and $\delta(s-s_b)$ denotes the one-hot representation of a sample drawn from a categorical distribution with PMF $p$. Jang et al. in~\cite{jang2016categorical} proposed a Gumbel-Softmax reparametrization trick to approximate samples from a categorical distribution with a differentiable function. We use this idea and replace $\delta(s - s_b),\, s_b \sim p$ with a sample from the Gumbel-Softmax distribution, i.e. 
\begin{align}
    \mathcal{L}_G(x, p) \approx \sum\limits_{b=1}^B \sum\limits_{s=0}^{d-1} q_{s, b} \, \mathcal{D}_{\phi}(\mathcal{M}_{s} x + \varepsilon_b)
\end{align}
\vspace{-5pt}
where 
\begin{align}
    q_{s, b} = \frac{\exp{((g_{b,s} + \log(p[s]))/\tau)}}{\sum\limits_{i=0}^{d-1} \exp{((g_{b, i} + \log(p[i]))/\tau)}}, \, g_{b,s} \sim \textrm{Gumbel}(0, 1).
    \label{eq:gumbel_approx}
\end{align}
Note that~\eqref{eq:gumbel_approx} is a continuous approximation of the $\arg \max$ function, $\tau$ is the softmax temperature factor and $q_{s, b} \rightarrow \delta(s - \arg \max_s{(g_{b,s} + \log p[s])})$ as $\tau \rightarrow 0$. Note that drawing samples from $\arg\max_s{(g_{b, s} + \log p[s])}$, $g_{b, s} \sim \textrm{Gumbel}(0,1)$ is an efficient way of sampling from $p$ distribution~\cite{jang2016categorical}. Furthermore, to obtain samples from the Gumbel distribution~\cite{madison2014sampling}, it suffices to transform samples from a uniform distribution using $g = -\log(-\log(u))$, $u \sim \textrm{Unif}(0, 1)$.

\vspace{-7pt}
\section{Implementation details}
\vspace{-6pt}
We present the pseudo-code for MSR-GAN in Alg.~\ref{alg:msrgan}. In all our experiments, we use a batch-size of $B=200$ and keep the number of real measurements as $N = 3\times 10^4$ unless otherwise mentioned. We have three separate learning rates for the discriminator, the signal and the PMF denoted by $\alpha_{\phi}$, $\alpha_x$ and $\alpha_{p}$, while in most experimental settings we keep  $\alpha_x\!=\!\alpha_p$. We reduce the learning rates by a factor of $0.9$, with different schedules for different learning rates. We use \texttt{SGD}~\cite{bottou2010sgd} as the optimizer for the discriminator and the signal $x$ with a momentum of $0.9$. We also update $p$ using gradient descent steps after normalizing the corresponding gradients. We clip the gradients of the discriminator to have norm $1$. Similar to common practice, we train the discriminator $n_{\textrm{disc}}=4$ times per updates of $x$ and $p$. To have stabilized updates with respect to $p$, we choose $\tau=0.5$ in our experiments. We also use spectral normalization to stabilize the training~\cite{miyato2018spectral}.

Our architecture of the discriminator consists of three fully connected (FC) layers with $\ell$, $\ell/2$, and $1$ output sizes, where $\ell$ is determined accordingly for different experiments.
We use $\texttt{ReLU}$~\cite{xu2015empirical} for the non-linear activations between the FC layers. We initialize the layers with weights drawn from normal distribution with mean zero and $0.01$ standard deviation. We train MSR-GAN for $30,000$ and $50,000$ iterations for high and low SNR regimes, respectively. To enforce $p$ to have non-negative values while adding up to one, we set it to be the output of a $\texttt{Softmax}$ layer. Our implementation is in PyTorch and runs on single GPU.
\vspace{-5pt}

\section{Numerical results}
\label{sec:results}
\vspace{-6pt}
\begin{figure}
\centering
	\begin{tikzpicture}
		\begin{groupplot}[group style={group size= 1 by 1,                      
    				horizontal sep=0.6cm, vertical sep=2cm},     
	            	 legend pos= south west,        
					 legend style={at={(0.65,0.2)}, legend cell align=left, draw=none, fill=none},
					 grid=both,                         
    				 height=4cm,width=9cm,
    				 xmin=1,xmax=60,
					 ymin=-0.1,ymax=1, 
					 ylabel near ticks, xlabel style={align=center,text width=3cm},
					 yticklabel style = {font=\footnotesize,yshift=0.5ex},
					 xticklabel style = {font=\footnotesize,xshift=0.5ex},] 
		\nextgroupplot[xlabel={Segment length ($m$)},ylabel={Success rate}, ymin=0,ymax=1.1, x label style={at={(axis description cs:0.5,-0.2)},anchor=north}] 	
			 \addplot[darkblue,very thick] table[x=x, y=y, col sep=comma]{figs/exp_mask_dat/mask_gan.dat};
			 \addplot[darkred,very thick] table[x=x, y=y, col sep=comma]{figs/exp_mask_dat/mask_manopt.dat};
			 \addplot[darkgreen,very thick] table[x=x, y=y, col sep=comma]{figs/exp_mask_dat/mask_em.dat};
		\legend{{\footnotesize MSR-GAN },{\footnotesize MSR-SIF } ,{\footnotesize EM}};
		\end{groupplot}
	\end{tikzpicture}
\caption{Effect of segment length on the success rate of 1) MSR-GAN (blue curve), 2) MSR-SIF (red curve), 3) EM (green curve). In this experiment the signal length $d=60$, the signal is generated randomly and $\sigma=0.01$. The success rate is computed based on $10$ random initializations for each segment length value. All three methods are initialized with the same random $x$ and $p$.}
\vspace{-10pt}
\label{fig:mask_exp}
\end{figure}
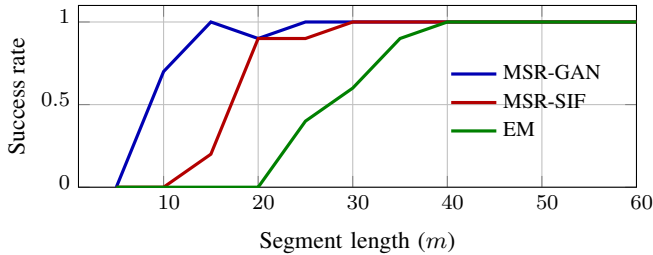

In this section we first provide details on our evaluation metrics and baselines. Next, we discuss our results.

\noindent \textbf{Evaluation metrics and baselines}: The $\textrm{SNR}$ of the observations is defined as the variance of the clean measurements divided by the variance of the noise. As the signal and PMF are reconstructed up to a random global shift, we align the reconstructions before comparing them to the ground truths. We use relative error (rel-error) between the aligned estimated signal $\widehat{x}$ and the ground truth $x$ as the quantitative measure of the performance, defined as $\textrm{rel-error} = \frac{\min_s \Vert x - \mathcal{R}_s \widehat{x} \Vert^2}{\Vert x \Vert^2}$, where $\mathcal{R}_s$ shifts its input by $s \in \{0, ..., d-1\}$. To assess the quality of the estimated PMF, we use total variation (TV) distance, defined as $\textrm{TV} = \frac{1}{2} \min_s \Vert p - \mathcal{R}_s \widehat{p} \Vert_1$~\cite{Chen2014tv}. We also define success rate by running MSR solutions with $10$ different initializations. The ratio of the initializations that lead to a relative-error less than a threshold $0.02$ is reported as the success-rate. 

We compare MSR-GAN to two baselines: 1) Estimating shift-invariant features, i.e. moments up to the third order, from the measurements and recovering $x$ and $p$ by solving a non-convex optimization problem~\cite{zehni_msr}. We use up to third order moments as the features. We call this baseline MSR via shift-invariant features (MSR-SIF). We use Riemannian trust-regions method~\cite{rtr} implemented in Manopt~\cite{manopt} to solve the optimization problem. 2) Expectation maximization (EM). In this baseline, we formulate MSR as a maximum marginalized likelihood estimation problem and solve it via EM~\cite{Bendory2017, zehni_msr}.

\vspace{5pt}
\noindent \textbf{Effect of knowledge of PMF on the MSR-GAN results}: Figure~\ref{fig:signal_viz_results} shows the results of MSR-GAN on different signals with $d=64$ and $m=24$ in three different scenarios: 1) $p$ in known (first column), 2) $p$ is not known but fixed with a uniform distribution during training (second column), 3) $p$ is not known and we recover it along side $x$ (third and fourth columns). Note that for all three scenarios, we are using Alg.~\ref{alg:msrgan} with the same discriminator architecture and $\ell=100$. However, for the first and the second scenarios, we do not update $p$ (skip step $9$-update $p$), rather keep it fixed with the true and the uniform distribution, respectively.

\begin{figure}
\centering
	\begin{tikzpicture}
		\begin{groupplot}[group style={group size= 1 by 1,                      
    				horizontal sep=0.6cm, vertical sep=2cm},     
	            	 legend pos= south west,        
					 legend style={at={(0.005,0.03)}, legend cell align=left, draw=none, fill=none},
					 grid=both,                         
    				 height=4.5cm,width=9cm,
    				 xmin=-0.5,xmax=3,
					 ymin=-3.5,ymax=0.5, 
					 ylabel near ticks, xlabel near ticks, xlabel style={align=center,text width=3cm},
					 yticklabel style = {font=\footnotesize,yshift=0.5ex},
					 xticklabel style = {font=\footnotesize,xshift=0.5ex},] 
		\nextgroupplot[xlabel={$\textrm{log}_{10} \textrm{SNR}$},ylabel={Relative-error},x label style={at={(axis description cs:0.5,-0.1)},anchor=north}] 	
			 \addplot[darkblue,very thick] table[x=x, y=y, col sep=comma]{figs/exp_noise_dat/noise_gan_m18.dat};
            
			 \addplot[darkred,very thick] table[x=x, y=y, col sep=comma]{figs/exp_noise_dat/noise_manopt_m18.dat};

			 \addplot[darkgreen,very thick] table[x=x, y=y, col sep=comma]{figs/exp_noise_dat/noise_em_m18.dat};
		\legend{{\footnotesize MSR-GAN },{\footnotesize MSR-SIF}, {\footnotesize EM}};

		\end{groupplot}
	\end{tikzpicture}
\caption{Comparison between MSR-GAN with different baselines in terms of relative-error versus SNR of the observations. In this experiment $d=60$ and $m=18$. All three methods have been initialized with the same signal and PMF and the reported results are the median across $10$ different initializations and noise realizations for the observations.}
\vspace{-10pt}
\label{fig:noise_exp}
\end{figure}
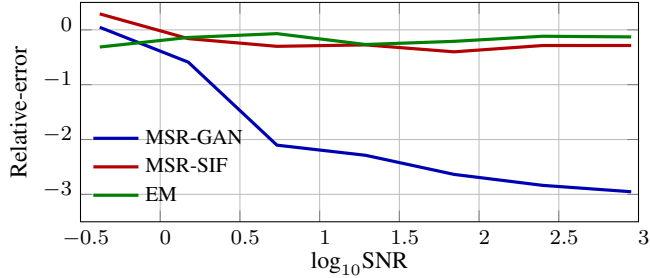

Note that when the PMF is known, the results of MSR-GAN closely match the ground truth signal. When the PMF is unknown, if we fix $p$ to be a uniform distribution (see second column of Fig.~\ref{fig:signal_viz_results}), we observe that although the reconstructed signal is close to the GT, it has larger relative error compared to the scenarios where the PMF is given (see the first column of Fig.~\ref{fig:signal_viz_results}) or the PMF is updated jointly with the signal (see the third column of Fig.~\ref{fig:signal_viz_results}).
Updating $p$ jointly with $x$ (Fig.~\ref{fig:signal_viz_results}-third column) leads to more accurate reconstruction of the signal. 
This shows the importance of recovering the distribution of the segments.  

\vspace{5pt}
\noindent \textbf{Effect of segment length and comparison with baselines}:
Figure~\ref{fig:mask_exp} illustrates the effect of segment length $m$ on the success rate of MSR-GAN compared to the other two baselines. For this experiment, we set the network hyper-parameter $\ell \!= \!300$ and test the performance of our algorithm on randomly generated signals of length $d = 60$. As discussed in~\cite{zehni_msr}, solving MSR for smaller segment length regimes using shift invariant features is more challenging, as the number of equations provided by the moments for smaller segment lengths can be less than the number of unknowns. Similarly, the EM algorithm fails at shorter segments, i.e. $m \leq 25$, where the success rate is less than $50\%$. EM is more likely to get stuck at a local optimal solution when the segment length becomes smaller. However, as MSR-GAN solves the inverse problem by matching the distribution of real measurements and stochastic gradient descent, it achieves higher success rates for smaller segment lengths. In particular, even at $m = 15$, MSR-GAN achieves a success rate close to $100\%$.

\vspace{5pt}
\noindent \textbf{Effect of noise and comparison with baselines}: In Fig.~\ref{fig:noise_exp}, we investigate the effect of noise on the performance of MSR-GAN compared to the baselines. 
For this experiment $d=60$, $m=18$ and for the discriminator's architecture we set $\ell=300$. 
Note that in different noise regimes MSR-GAN outperforms MSR-SIF and EM. Here we have a short segment length, thus as mentioned earlier solving MSR is more challenging and both baselines get stuck in local minima that is not close to the ground truth solution. Note that if we increase the segment length we observe an improved reconstruction error and success rate for MSR-SIF and EM (as also observed in Fig.~\ref{fig:mask_exp}). This suggests that MSR-GAN is a better solution compared to the baselines in short segment length regimes.



\section{Conclusion}
\label{sec:conclusion}
\vspace{-6pt}
In this paper, we focused on the multi-segment reconstruction (MSR) problem, where we are given noisy randomly located segments of an unknown signal and the goal is to recover the signal and the distribution of the segments. We proposed a novel adversarial learning based approach to solve MSR. Our approach relies on distribution matching between the real measurements and the ones generated by the estimated signal and segment distribution. We formulated our problem in a Wasserstein GAN based framework. We showed how the generator loss term is a non-differentiable function of the segments distribution. To facilitate updates of the distribution through its gradients, we approximate the loss function at the generator side using Gumbel-Softmax reparametrization trick. This allowed us to update both the signal and the segment distribution using stochastic gradient descent. Our simulation results and comparisons to various baselines verified the ability of our approach in accurately solving MSR in various noise regimes and segment lengths. 

\clearpage
\newpage
\bibliographystyle{IEEEbib}
\bibliography{references.bib}

\end{document}